\documentclass[twocolumn,aps,pre,a4paper,showpacs,10pt,flushbottom,final]{revtex4-1}
\usepackage{amsbsy}
\usepackage{amssymb}
\usepackage[dvips]{graphicx}

\begin{document}
\noindent LPT Orsay 11-23

\title{Motifs emerge from function in model gene regulatory networks}

\author{Z. Burda $^1$, A. Krzywicki $^2$, 
and O.C. Martin $^{3,4}$ and M. Zagorski $^1$} 

\affiliation{$^1$ Marian Smoluchowski Institute of Physics
and Mark Kac Complex Systems Research Centre, Jagellonian University,
Reymonta 4, 30-059 Krakow, Poland\\
$^2$ Univ Paris-Sud, LPT ; CNRS,
UMR8627, Orsay, F-91405, France.\\
$^3$ Univ Paris-Sud, LPTMS ; CNRS, UMR 8626, 
F-91405, Orsay, France\\ 
$^4$ INRA, CNRS, UMR 0320 / UMR 8120 G\'en\'etique V\'eg\'etale, 
F-91190 Gif-sur-Yvette, France}

\date{\today}

\begin{abstract}
Gene regulatory networks 
arise in all living cells, allowing the control of
gene expression patterns. The study of their topology has revealed
that certain subgraphs of interactions or ``motifs''
appear at anomalously high frequencies. We ask here whether
this phenomenon may emerge because of the functions carried out by
these networks. Given a framework for describing
regulatory interactions and dynamics, we consider in the
space of all regulatory networks those that have a prescribed function.
Monte Carlo sampling is then used to determine how these functional networks 
lead to specific motif statistics in the interactions. 
In the case where the regulatory networks are constrained to exhibit
multi-stability, we find a high frequency of gene pairs
that are mutually inhibitory and self-activating. In contrast,
networks constrained to have periodic gene expression patterns (mimicking
for instance the cell cycle)
have a high frequency of bifan-like motifs involving four genes
with at least one activating and one inhibitory interaction.
\end{abstract}

\vspace{2pc}

\maketitle

\vspace{12mm}



\section{Introduction}
Both natural and artificial networks have unexpected properties
that may find their origin in the way they were constructed. However,
another possibility, in particular in the context of biological networks,
is that constraints associated with network \emph{functionality}
are the main determinants of these unexpected properties.
We focus here on gene regulatory networks (GRN), the set of 
interactions between genes as well as the rules
for expression dynamics
that allow all living cells to control gene their expression patterns.
In the last decade, gene interactions have been measured, modified, 
engineered, \emph{etc.}, and 
so quite a lot is known about how any given gene can
affect another's expression. Furthermore, small 
gene networks have been designed
to implement simple functions
\emph{in vivo}~\cite{Elowitz_Leibler_2000,Gardner_Cantor_2000}, 
and much larger sets of interactions have been reconstructed in a number of 
organisms~\cite{HerrgardCovert2004,SalgadoGama2006,Hu_Killion_2007}.
From these large networks it has been possible to show that
several ``motifs'' -- subgraphs with given interactions --
arise far more often than might be 
expected~\cite{
Shen-Orr_Milo_2002,Ma_Kumar_2004,Lee_Rinaldi_2002,ZhuZhang2008}.
One of the
most studied motif is the so called \emph{Feed
Forward Loop} or FFL, a graph based on three genes 
where the first regulates the second, and both the first and the
second regulate the third. Another example is the bifan motif
in which 2 genes control two others.
Biological functions have
been proposed for these motifs~\cite{Mangan_Alon_2003,Camas_Poyatos_2008} 
which give them some meaning, but one may ask whether other motifs could
perform the same functions and what level of enrichment 
might be expected if function were the sole cause of 
motif over-representation. Unfortunately,
the functions of GRN and the constraints they must satisfy
(e.g., kinetic response characteristics or robustness to noise)
are still poorly understood, so such questions cannot
be addressed in a truly realistic framework. Instead, we
will (i) work within a plausible model of transcriptional regulation,
(ii) impose functional constraints on the patterns of gene expression,
(iii) determine which motifs emerge when considering
the space of all possible functional GRN. 
This particular task is related to previous 
work that used genetic algorithms or simulated annealing
to design genetic networks having given functional 
properties~\cite{FrancoisHakim2004,FrancoisHakim2007,RodrigoCarrera2007}.
Those studies found that the optimization procedures indeed led
to particular architectures. Our approach differs by not relying
on a design procedure: we want 
to get away from any dependence on the optimization
algorithm and see how functionality \emph{on its own}
constrains the possible architectures. In this framework, two types
of constraints will be applied: we will impose either a set
of steady-state expression patterns, or a time periodic pattern
of expression motivated by previous studies of cell cycling.
Interestingly, we find very different motifs for these two cases;
in Alon's \cite{AlonBook2006}
terminology, bifan, diamond and four point cycle
motifs appear only in the second case.

Our model of transcriptional regulation is simple
enough to be used for illustration and, hopefully, for identification
of the generic features of genetic networks, but at the
same time is rooted
in bio-physical reality to avoid ad-hoc assumptions.
It significantly extends the framework
of ref. \cite{BKMZ_2010}, in particular by allowing for 
inhibitory interactions. We begin by describing our model and 
then show its properties, in particular the kinds of motifs that 
emerge from the functional constraints imposed on the networks.

\section{The model}

\subsection{Transcription factor binding}
We start with $N$ genes coding for transcription factors
that may influence each other's expression.
To keep the model as realistic as possible, 
we include the known biophysical determinants of 
transcriptional control. In particular, the binding of a
transcription factor (TF) to a site is described
thermodynamically~\cite{HippelBerg1986,BergHippel1987,GerlandMoroz2002}
and depends on the mismatch of two character strings
of length $L$,
one for the TF and one for the binding site. 
Up to an additive constant, the associated free energy 
in units of $k_B T$ is taken to be
$\varepsilon d_{ij}$ where $d_{ij}$ is the number of mismatches,
$T$ is the temperature and $k_B$ is Boltzmann's constant.
The parameter $\varepsilon$ is a penalty per
mismatch which has been measured
experimentally to be between one and three
if each base pair of the DNA is represented by one
character~\cite{SaraiTakeda1989,StormoFields1998,BulykJohnson2002}.
Also, by comparing to the typical number of base pairs 
found for experimentally studied binding sites, one has
$10 \le L \le 15$. 
For all the work presented here, we use
$\varepsilon=2$ and $L=12$, but we have checked that our conclusions 
are not specific to these values.

We shall define the ``interaction
strength'' $W_{ij}$ from gene $j$
to gene $i$ via the Boltzmann factor
\begin{equation}
W_{ij} = e^{-\varepsilon d_{ij}} / Z  
\label{eq:Wij}
\end{equation}
where $Z$ is normalization (a partition function).
If there were just one transcription factor molecule of type $j$, 
$W_{ij}$ would be the probability to find that molecule bound
in gene $i$'s regulatory region.
Gerland et al.~\cite{GerlandMoroz2002} have shown
that in practice $Z$ in Eq.~(\ref{eq:Wij}) is 
close to 1 and that the probability of finding any
given TF molecule bound rather than unbound is quite low.

\begin{figure}
\centering
\includegraphics[width=8cm]{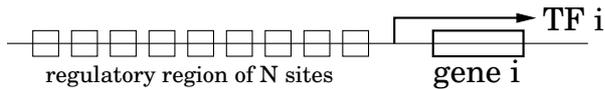}
\caption{\label{fig:model} Each gene's regulatory region
contains $N$ binding sites, one for each of the $N$
transcription factors produced by the $N$ genes.
The probability of occupation (POCC~\cite{Granek_Clarke_2005}) of the
regulatory region 
determines the average transcription rate of the gene $i$ 
under consideration.}
\end{figure}

For simplicity, 
and to prevent different TFs from accessing a same site,
we use a standardized form of regulatory region for each
gene. This situation is illustrated in Fig.~\ref{fig:model}
for gene $i$ which produces the transcription factor $TF_i$.
The regulatory region of
each gene has $N$ binding sites,
one dedicated to each of the $N$ different TF types. 
Suppose, that there are $n_j$ TF molecules of type $j$ that can bind to 
the site $j$ in gene $i$'s regulatory region; given 
that this site can be occupied
by only one TF molecule at a time, it is necessary to take into account
possible competition effects between all molecules $n_j$
of type $TF_j$. Using the fact that $Z$ in
Eq.~(\ref{eq:Wij}) is close to 1, it is possible to approximate
the occupation probability of the binding
site by~\cite{GerlandMoroz2002}:
\begin{equation}
P_{ij} = \frac{1}{1 + 1/(n_j W_{ij})}  \; .
\label{eq:probOccupy}
\end{equation}
As emphasized in \cite{BKMZ_2010}, 
$P_{ij}$ depends strongly on $d_{ij}$ and is 
appreciable only when the mismatch is small, which is an 
a priori unprobable event imposed in functional genotypes by 
the selection pressure. 

\subsection{Transcriptional control}
Again for pedagogical reasons, we shall consider that all genes have
the same maximal transcription rate; denoting
by $n$ the associated maximum number of 
TF molecules in the system of a given type, we shall set
$n_j = S_j n$ where $S_j$ is then the current
(normalized) level of transcription for gene $j$, ranging
between 0 and 1.
Experimentally, $n$ is known to range from 
of order unity to many
thousands~\cite{GoldingPaulsson2005,BecskeiKaufmann2005,ElfLi2007}.
Here we shall use $n=1000$, but again we have checked that
using values ten times smaller or larger does not change
our conclusions.

The expression $S_i$ of gene $i$ will vary with
the presence of transcription factors bound in
its regulatory region, but present knowledge
does not provide us with quantitative information on this
dependence. Much past modeling 
work~\cite{KauffmanBook1993,Wagner1996,
BornholdtRohlf2000,LiLong2004,AzevedoLohaus2006}
has dealt with this
obstacle by considering that each occupied binding site 
provides an activating or inhibitory signal and that
all signals are then added and compared to a threshold:
below (respectively above) this threshold, transcription
is off (respectively on).
However, more recent experimental work
and associated modeling~\cite{Segal_Widom_2009,Giorgetti_Siggers_2010}
suggests that transcription rates
\emph{in vivo} can exhibit graded responses. 
This result is not surprising given that transcription
factors are sometimes bound and sometimes not, so any
\emph{average} transcription rate has no reason to be
binary. Our work thus follows~\cite{Segal_Widom_2009,Giorgetti_Siggers_2010,BKMZ_2010}
by considering continuous transcription rates determined by
the probabilities that binding sites are occupied.

Consider first the case where all $N$ transcription factors affect gene $i$
as activators. If at least one of the binding sites is occupied by its
TF, we consider that the gene will be transcribed; this choice corresponds to
having the transcription rate be proportional to
the Probability of OCCupation or ``POCC'' \cite{Granek_Clarke_2005} 
of the regulatory region. 
Calling $P_i$ this probability, we have
\begin{equation}
P_i = \sum_{k=1} \sum_{[j_1,...,j_k]} P_i^{(k,N-k)}(j_1,...,j_k,
\bar{j}_{k+1},...,\bar{j}_{N})
\label{Pi}
\end{equation}
where $j_l$ ($\bar{j}_l)$ is the label of an occupied (unoccupied) 
binding site and $[j_1,...,j_k]$ stands
for a combination of $k$ out of $N$ gene labels.
Assume now that the bindings arise independently (no cooperativity),
{\em i.e.} that the probabilities in the sum  
factorize into a product of terms $P_i^{(1,0)}(j)$ 
(or $P_i^{(0,1)}(j)$). Replacing
$P_i^{(1,0)}(j)$ by $P_{ij}$ defined 
in Eq.~\ref{eq:probOccupy} we write
\begin{equation}
P_i^{(k)}(j_1,...,j_k,\bar{j}_{k+1},...,\bar{j}_N) =
\prod_{j} P_{ij} \times \prod_{j'} (1 - P_{ij'})
\label{Piupk}
\end{equation}
where $j$ runs over indices for which there is binding and $j'$
runs over the other indices.
Now, the sum over $k$ in Eq. \ref{Pi} can be explicitly 
performed. In addition, we identify
up to an overall scale the transcription
rate with the probability of occupancy $P_i$. Considering
that protein content is proportional to transcription rate (at least in the
steady state), we set $S_i$ -- the mean normalized
expression level of gene $i$ -- equal to $P_i$. One finally gets 
\begin{equation}
S_i = 1 - \prod_j (1 - P_{ij})   
\label{eq:activators}
\end{equation}
which is the basic equation of the ``mean field'' 
model of ref.~\cite{BKMZ_2010}.
The neglect of fluctuations and the corresponding limitations 
were already explained there.
%
Note that if the $P_{ij}$ are small, transcription is additive
in these variables, while in the binary limit where
$P_{ij}$ is 0 or 1, $S_i$ corresponds to the logic
of transcription being ``on'' if and only if at least one of the 
binding sites is occupied, as expected from the
use of the POCC.

Our treatment of \emph{inhibitory} interactions 
(due to repressors) is new and is motivated
by a number of known cases where the binding of a TF
acts as a veto. One way this can happen is if the presence of the TF makes
the DNA form a loop that conceals
the other binding sites.
Another mechanism for vetoing transcription is simply for the bound TF to block
the advance of the polymerase.
Within our framework, transcription proceeds as in Eq.~\ref{eq:activators},
in the absence of such repressors bound
to their sites, 
but as soon as any of the inhibitory sites are bound,
transcription is turned off. Assuming cooperative
effects are absent as before,
and repeating for repressors the argument just used for activators,
we are led to modify Eq.~\ref{eq:activators} to
\begin{equation}
S_i = [1 - \prod_{j} (1 - P_{ij}) ] \prod_{j'} (1 - P_{ij'})
\label{eq:inhibitors}
\end{equation}
where $j$ runs over activating interactions and $j'$ over inhibitory
interactions.

The transcriptional dynamics is then defined as follows.
Just like in many other modeling frameworks, we take
time to be discrete~\cite{KauffmanBook1993,Wagner1996,
BornholdtRohlf2000,LiLong2004,AzevedoLohaus2006}; 
at each time step we first update
the $P_{ij}$ in Eq.~\ref{eq:probOccupy}
(using $n_j = n S_j$) and then update the $S_i$ in Eq.~\ref{eq:inhibitors}.
These updates are deterministic, and in general the system goes towards a
fixed point (corresponding to steady-state expression levels)
or towards a cycle (corresponding to periodic behavior of the expressions
in time).

\par
By neglecting cooperative effects, we obtain a toy model where the only
parameters are those determining the binding probabilities implicit
in Eq.~\ref{eq:probOccupy}
and these are subject to experimental constraints. Incorporating
cooperative effects could lead to a more realistic model
but at the cost of more parameters.
For instance, one could replace in 
Eq. \ref{Piupk} the equality by a proportionality.
Such an assumption often appears in the literature: using the stationary 
limit of appropriate kinetic equations, one argues that the concentration of a
molecular complex is proportional to the product of concentrations of the
constituents. Here, because of the reparametrization symmetry of the
dynamics, the proportionality constant can only depend on $k$. One could then
truncate the sum over $k$, say at $k=3$, to avoid too many
free parameters, a situation that arises in a number of genetic network 
reverse engineering attempts. 
Such a model deserves study, but this 
is beyond the scope of the present work.

\subsection{Genotypes and Phenotypes}
As previously mentioned, the TFs and their binding sites are
associated with character strings. We are interested in
the space of all GRN, which means here all possible character
strings.
However, it is easy to see 
that all choices of TF character strings
are equivalent, so we can fix them without any loss of generality.
(Biologically, it is known that TF and most protein
coding genes are far more conserved than the TF binding sites.
See ref. \cite{BKMZ_2010} for a discussion of this point.)
Any given GRN is then completely specified by the $N^2$ character 
strings of its binding sites and by the 
specification of the activating or inhibiting
nature of each interaction. Since DNA bases come in four types,
A,C,G,T, we use an alphabet of four characters for our strings.
This set of strings is referred to as the ``genotype'' of the GRN.
Clearly the most relevant quantities in a genotype are
the mismatches $d_{ij}$ of these $N^2$ strings to their TF string.
A genotype can then usefully be represented by this $N$ by $N$ matrix
of mismatches or by the corresponding 
matrix of interaction strengths $W_{ij}$, plus the sign
(activating vs. inhibitory) associated with each of these interactions.

At any time step $t$, the pattern of mean gene expression
can be represented by the 
vector ${\mathbf S} (t) = \{ S_j (t) \}_{j=1,\ldots N}$.
We shall consider two classes of functions to be imposed
on our GRN. The first
is motivated by cell types in multi-cellular organisms: 
we want the GRN to be able to have steady state expression
vectors that are very close to 2, 3, or more \emph{target} patterns,
each associated with a different tissue.
Note that some such patterns involving a dozen or so genes 
have been inferred in various
organisms~\cite{Espinosa-Soto_Padilla-Longoria_2004,Chickarmane_Peterson_2008}.
The second kind of function we shall impose is for the vector
to follow tightly and step by step a sequence of 
patterns that forms a target \emph{cycle}. Such cases of cycling GRN
have been studied previously within threshold and boolean 
models~\cite{LiLong2004,Davidich_Bornholdt_2008}. 

For each type of functional constraint imposed, we refer to the
``phenotype'' of the GRN as (i) the different steady-state expression
vectors for the first case; (ii) the cyclic pattern of expression
vectors for the second case.
Given a GRN genotype, determining its phenotype is straightforward
in practice. In the first case where we have
given target expression patterns, we start in these target vectors
and we see whether we converge to a nearby fixed point under 
iteration of the transcriptional dynamics. 
(In contrast, in our previous work, we had considered initial states
that were unrelated to the target vector.)
In the second case,
we start with one of the patterns in the target cycle and
see whether the trajectory under iterations stays close to that cycle.
For the steady state behavior, 
we shall impose 2, 3, or more vectors that consist of
$N/2$ levels at 0 and $N/2$ at 1, and furthermore that are
taken to be orthogonal (for the 0/1 coding for $S_i$ this means that
the scalar product of two vectors is $N/4$). Setting $N=16$ (the 
choice of $N$ is not important as long as it has a moderate value, but we
have not explored what happens at large $N$),
we define four mutually orthogonal targets as follows,
a direct generalization of that of ref. \cite{BKMZ_2010}:
\par\noindent
1 1 1 1 1 1 1 1 0 0 0 0 0 0 0 0\\
1 1 1 1 0 0 0 0 1 1 1 1 0 0 0 0\\
1 1 0 0 1 1 0 0 1 1 0 0 1 1 0 0 \\
1 0 1 0 1 0 1 0 1 0 1 0 1 0 1 0 \\
This choice is motivated by the fact
that at large $N$, random binary vectors are 
typically nearly orthogonal. The symmetries of the model are
relevant for studying it, but with any interesting choice
of vectors, most of the symmetries are broken.
Note that since one can transform
these vectors into each other by index permutation, the basins of
attraction leading to these targets are on average of equal size.

For the case where one enforces a target cycle, we shall use the toy sequence
where the genes are taken to lie on a ring, and
the cycle consists in having the ``on'' genes shift to the right
at each time step:
\par\noindent
1 1 1 1 1 1 1 1 0 0 0 0 0 0 0 0 \\
0 0 1 1 1 1 1 1 1 1 0 0 0 0 0 0 \\
0 0 0 0 1 1 1 1 1 1 1 1 0 0 0 0 \\
0 0 0 0 0 0 1 1 1 1 1 1 1 1 0 0 \\
0 0 0 0 0 0 0 0 1 1 1 1 1 1 1 1 \\
1 1 0 0 0 0 0 0 0 0 1 1 1 1 1 1 \\
1 1 1 1 0 0 0 0 0 0 0 0 1 1 1 1 \\
1 1 1 1 1 1 0 0 0 0 0 0 0 0 1 1 \\
1 1 1 1 1 1 1 1 0 0 0 0 0 0 0 0 \\
This is reminiscent of the cycle studied
by Li \emph{et al.}~\cite{LiLong2004}
for the yeast cell cycle. We have also considered a
cycle where the shift is not by two, but by 1 or 3 steps. The
results are nearly the same in all these cases.

To quantify the deviations from the ideal
target behavior, we first check whether we have
steady-state behavior (in the first case) or cyclic behavior
(in the second). 
For each target vector ${\mathbf S^{(target)}}$, we 
define its distance to the associated GRN specific expression vector 
${\mathbf S}$ via:
\begin{equation}
D({\mathbf S}, {\mathbf S^{(target)}})=\sum_i \mid S_i-S_i^{(target)}\mid \; .
\label{eq:distance}
\end{equation}
By summing all these distances, one for each target
(each steady state in the first case, and
each expression vector of the periodic cycle in the second), we
obtain what we refer to as the ``total'' distance $D_T$ for that GRN.
The resulting measure of ``fidelity'' to the
imposed function can be turned into a kind of fitness via
\begin{equation} 
F(GRN)=\exp{(-f D_T)}
\label{eq:fitness}
\end{equation} 
where $f$ acts as a control parameter allowing one
to be more or less stringent on the fidelity.
We thus consider the set of all GRN and apply
the relative weight $F(GRN)$ to each; this
then provides an \emph{ensemble} for the GRN, and by 
adjusting $f$ we can focus
on those GRN that are the most functional. 
For specificity, we shall work with
$f=20$, but our results depend only very weakly 
on this choice provided $f$ is in the range 10 to 100.

\subsection{MCMC sampling}
\begin{figure}
\centering
\includegraphics[width=8cm]{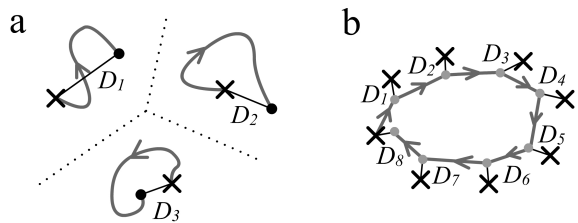}
\caption{\label{fig:MCMC} A schematic representation of our MCMC
process. (a) Steady state behavior and $n=3$: crosses (heavy dots)
stand for the target (fixed  point) states, while the line is
the system's trajectory. The "total" distance entering the 
Metropolis test is $D_T=D_1+D_2+D_3$. (b) Similar as before,
but for a cycle. Grey dots stand for successive states obtained by
iterating Eq. \ref{eq:inhibitors}. Here $D_T=D_1+ \dots + D_8$. 
}
\end{figure}

To sample our ensemble of constrained genotypes, we apply
a Monte Carlo Markov Chain (MCMC) using the Metropolis rule.
This computer algorithm produces a (biased) random walk in
the fitness landscape that visits at long times
the different genotypes according to their fitness
as given in Eq. \ref{eq:fitness}; the sampling thus focuses
on genotypes having high fidelity to the imposed functions.
In detail, we perform random mutations of
the binding sites. This produces changes to the
edge weights and thus to the genotypes. (Technically,
it would be possible to work at the level of edge weights
alone, but it would make explanations of the MCMC far
more delicate.)
A {\em sweep} is defined as $LN^2$ successively
attempted  changes of the genotype (a random mutation of one coding 
letter and, independently, a random switch of 
the sign of one of the TF-DNA interactions). Each such change is  
accepted or rejected by the Metropolis algorithm. 
We always make a hot start, using as input a completely
random GRN. After some time, as in ref. \cite{BKMZ_2010}, we produce
a GRN sufficiently close to the target (see Fig. 
\ref{fig:MCMC}), and we use this to start
the production run of the MCMC. Thereafter, we iterate sweeps,
recording successive GRNs. Unfortunately, the simulation 
requires a lot of computation time, especially
for $L=12$, where small mismatches become very unprobable. Therefore,
we resort to the following modified procedure. We first set $L=8$ 
and generate an MCMC sampling, recording genotypes every 100 sweeps.
Since our dynamics depends on mismatches and not on the value of $L$, these
recorded GRNs are also fit at $L=12$, except that the distribution
of the magnitudes of the mismatches is wrong. Hence we set $L=12$ and 
upgrade our GRNs, obtaining in this way a sample of fairly independent
genotypes in a reasonable time. 

\subsection{Essential interactions and the essential network}
As already mentioned, genotypes can be represented by the $N$ by $N$ matrix
of entries $W_{ij}$s along with the $N^2$ signs specifying
the activating \emph{vs.} inhibitory types of each interaction.
These $W_{ij}$ are never
zero (\emph{cf.} Eq.~\ref{eq:Wij}), so we cannot say
that an interaction is completely absent.
Nevertheless, one may expect some interactions to be
more important than others, for instance when the $W_{ij}$s
are larger than average. An arbitrary cut-off could
be introduced for separating small and large values, but
it is better to base such a classification on functionality.
We thus consider what happens when an interaction $W_{ij}$
is removed by setting it to zero. Starting with one of the
genotypes generated by our MCMC (and thus
typically satisfying well the soft functional constraints),
we determine the change in fitness produced by
setting $W_{ij}$ to zero: if the change is rejected by Metropolis
in five successive attempts, we say that this
interaction is \emph{essential}, motivated by
the corresponding biological definition (a very
similar result is obtained by defining the
essentiality as the sensitivity to a single
deleterious mutation; since the definition of
essentiality involves the Metropolis test, a random
event, one sometimes finds false essentials, however
this is a very weak effect).
This definition leads to a \emph{summary}
description of a genotype via a list of pairs $(i,j)$ 
specifying the essential interactions as well as
their nature (activating or inhibitory). This
can then be represented by a directed graph, with
+ signs on the edges that are activating and - signs
on the edges that are inhibitory. Hereafter we refer to this
oriented and signed graph as the \emph{essential network} of the genotype;
note that no information on the weights of the interactions is attached
to this network representation.
\section{Results}
\subsection{Abundance of functional GRN}
The space of all GRN is finite in 
our framework since each genotype can be specified by 
$N^2$ character strings of length $L$ and the signs
of the associated interactions. In this space
we impose the soft constraint that a GRN implements
a function specified by a certain target expression behavior.
Is such a constraint very stringent? To find out,
we have generated millions
of random genotypes and find that none of them have 
good expression behavior: their fitness is orders of magnitude
lower than what we obtain from our MCMC sampling.
Thus, as in other gene network models~\cite{CilibertiMartin2007b},
by focusing on 
``functional'' GRN, we are considering an extremely small
subset of all GRN; these very rare GRN may thus very well be
atypical in many of their properties. Nevertheless, as long as
the number of constraints is not too high (the number
of steady states or length of the cycle cannot
grow indefinitely), the number of high fitness 
genotypes is huge. Indeed, our MCMC is able
to produce essentially as many different GRN as we want
even though the ensemble of interest is only an infinitesimal
part of the whole; this feature arises also in other
genotype to phenotype mapping models such as RNA neutral
networks~\cite{WagnerBook2005}.

We have also checked that the basin of attraction of a target phenotype
represents a large fraction of all possible
initial phenotypes, so that the fact of performing the imposed
function is not merely a dynamical accident. These basins
constitute approximately 99.8(9)\%, 52.9(9)\% and 49.6(8)\% of the whole
space for $n=$ 2, 3 and 4 respectively. As already explained, 
with our choice of target phenotypes, the basins associated with
individual targets are equal (after averaging over functional
GRNs).

\subsection{Functional GRN have sparse essential networks}
A question that comes to mind is whether
essential interactions are frequent or not. 
Consider first the case of the multi-stability phenotype
where we impose 1, 2, 3, ... steady states. In previous
work~\cite{BKMZ_2010} on a simpler model allowing no
inhibitory interactions, we found that
imposing a single steady state led to sparse essential interactions,
with the great majority of genotypes having just
one essential interaction $(i,j)$ for each gene $i$.
In the present model 
allowing for inhibitory interactions,
this property remains true (see Table~\ref{tab1}, where 
some of our results are summarized).
As we impose more steady states,
the mean number of essential interactions grows; again
each gene $i$ will typically have just a few essential interactions
(and almost never none),
with a mean of $1.2$, $1.5$, $1.9$ for 
2, 3, and 4 steady states at $N=16$. Furthermore,
these means are quite stable as one increases $N$,
so the essential networks of our functional genotypes form 
\emph{sparse} graphs. The mean degree of these networks
is insensitive to $N$, but grows when the number 
of constraints imposed on the GRN is increased.
\begin{center}
\begin{table}
\caption{\label{tab1} A sample of results of our simulations. nFP
stands for phenotypes with $n$ fixed points. The
robustness is defined as the frequency of genotypes surviving a
random mutation according to the Metropolis rule. 
Notice that it is fairly well reproduced
by $1-n/2N$, a result generalizing an analogous result of ref. 
\cite{BKMZ_2010}. The distance to the target is the distance
entering the fitness. It is nearly constant
when divided by the number of target phenotypes. Further
division by the number of active genes in a phenotype, 
{\em i.e.} $N/2$, yields approximately
6\%, which measures the average deviation of their activity 
from the maximum $S_i=1$.}
\begin{center}
\begin{tabular}{|l|l|l|l|l|} \hline \hline
observable & 2FP & 3FP & 4FP & cycle\\ \hline
 & & & &\\
$\langle \mbox{\rm \#essentials}\rangle$  & 14.70(1)
  & 21.08(2) & 29.39(1)  &23.42(1)  \\
$\langle \mbox{\rm \#repressors}\rangle$  & 2.500(12)
 & 6.38(2)  & 10.11(1)  &7.38(1)   \\
$\langle \mbox{\rm robustness}\rangle$    & 0.9494(2)
 & 0.9259(2)& 0.8998(3) & 0.9099(3)\\
$\langle \mbox{\rm dist2target}\rangle$   & 0.966(3)
  & 1.390(4) & 1.961(4)  &3.583(4)  \\
 & & & &\\
\hline \hline
\end{tabular}
\end{center}
\end{table}
\end{center}

The situation is similar when a cyclic phenotype is 
imposed: only a small fraction of the interactions turn out to be
essential. For the toy cases where the
genes are on a circle and the
cycle shifts the ``on'' genes by steps to the right,
we find again that the essential networks associated
with the genotypes of our MCMC ensemble are sparse,
and that the connectivity hardly changes as
one increases the number of genes. 

There is
a simple explanation for this sparseness:
at the level of the GRN, introducing an additional essential
interaction generally means increasing a $W_{ij}$. That 
has a high entropy cost as can be seen from 
the mismatches (there are few strings that have a low
mismatch, and many that have a high mismatch).
On the contrary, if one were to consider a Boolean model at the level 
of the essential network (to go from genotypes to phenotypes) and ignoring 
the molecular basis of the interactions, one would 
inevitably have far more functional graphs with 
dense interactions than with sparse interactions; sparseness would then
have to be enforced in an ad-hoc way since biological networks are
indeed sparse experimentally~\cite{ThieffryHuerta1998,BalazsiHeath2008}.

\subsection{Functional essential networks have 
parsimonious inhibitory interactions}
Are inhibitory interactions as frequent
as activating ones? The answer to this question depends on
the interactions considered. Indeed, even though 
the genotypes generated by the MCMC 
sampling have functional constraints, the many 
small $W_{ij}$ arising in genotypes have hardly any 
effect on the phenotype; their sign will thus
be random, and in effect they act like noise.
If instead we focus on the larger $W_{ij}$, 
the functional constraints are likely to bias 
the sign in favor of activating interactions.
To avoid an arbitrary definition of large
weights, we again use the notion of essentiality
because of its link with phenotypes. For 
the essential networks produced from the GRN of our MCMC
with the constraint of 2 to 4 steady states, 
we find that the great majority of the interactions
are activating, cf. Table~\ref{tab1}
(these numbers are not sensitive to $N$).
These results are not surprizing: increasing the number
of constraints forces the connections to be more complex
and to make greater use of inhibitory interactions. 
In the toy cases of genes
on a ring, we also see this general picture and find that
the number of both activating and inhibitory essential
interactions grows linearly with $N$.

\subsection{Abundance of functional essential networks}
Another question of interest concerns the number of distinct functional
\emph{essential networks} (the number of distinct GRNs is of 
little interest, being
trivially enormous since all inessential interactions can be changed
at will without affecting the phenotype). 
It is wise to first find the essential networks
that are in a sense representative for a group of GRNs, in other words
to perform a cluster analysis of the sample of essential networks 
at our disposal.
Let the numbers of such networks be $M$ and define a distance between a pair
of them, for example
\begin{equation}
\mbox{\rm Distance}(A,B) = \sum_{ij} (A_{ij} - B_{ij})^2
\end{equation}
where $A_{ij}$ (viz. $B_{ij}$) is $\pm 1$ for essential interactions and
0 otherwise. Our question can now be reformulated more precisely: does the
number of clusters, considered a proxy for the number
of representative essential networks, saturate at 
some moderate value as $M$ or not? (It must saturate somewhere, of course,
but that may be at very large values.)
To answer this, it is most convenient to use the modern 
affinity propagation algorithm
\cite{FreyDueck2007}, where the number of clusters is not preassigned but
is determined by the algorithm; the code can be downloaded from 
www.psi.toronto.edu/index.php?q=affinity propagation. 
As an illustration, for 4 fixed
points we find that the number of clusters grows at large $M$ roughly 
like $M^{2/3}$ (with a prefactor of the order of 0.3) and shows no 
sign of saturation up to at least $M=4000$. Other values of
$n$ lead to similar results, but some care is necessary 
in interpreting these trends at $n=2$ and $3$. 
Indeed, it turns out that for these 
values of $n$ many clusters, distinct
according to eq. (9), have essentially the same
topology and differ merely by the labeling of
nodes (this reflects symmetries in our choice 
of the target phenotypes). In contrast, for $n=4$
the clusters are genuinely different. To get more
insight into this problem, we have carried out a
complementary investigation, counting the number of
distinct topologies (instead of using the clustering
algorithm). This is very tedious and our account of
the network reparametrizations was only partial. With this proviso, it appears that
the number of distinct topologies again increases
like a power of M, however now the exponent
increases with $n$ (approximately from 0.69 for 
for $n=2$ to $0.97$ for $n=4$).

\begin{figure}
\centering
\includegraphics[width=8cm]{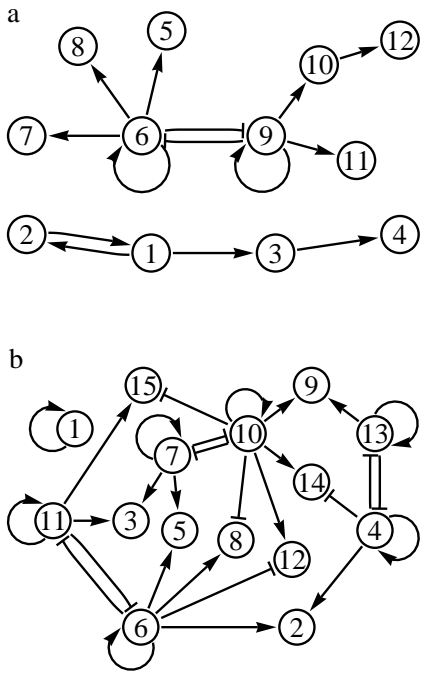}
\caption{\label{fig:circuits_nFP} 
The most common essential network topologies when $n$ steady-state expression
patterns are imposed. In each case, we see the 
presence of the motif with two mutually inhibiting and 
self-activating genes. Interactions shown are 
essential, and those genes whose target expression is the
same in all the steady states are omitted since they
provide no information. (Data for $N=16$; sub-figures a and b
are for $n=2$ and $4$. These topologies arise for over 20\% and
less than 1\% of the networks, respectively).}
\end{figure}

Beyond clusters, we can also ask
which essential network topologies are the most frequent. In 
Fig.~\ref{fig:circuits_nFP} we display the most frequent
topology when imposing $n=2$ and 4 fixed points. As $n$ increases,
the connections becomes more complex as expected; in particular at
small $n$ much of the topology is tree-like; to a large extent,
this reflects the sparseness and parsimony of these essential networks.
Note that our GRN are
connected by a succession of point mutations. This shows that
the same function can be performed by a very large number of distinct
networks, a feature also found in other models of 
gene networks~\cite{CilibertiMartin2007b}, but here we show
that many different \emph{topologies} arise too.
\begin{figure}
\centering
\includegraphics[width=8cm]{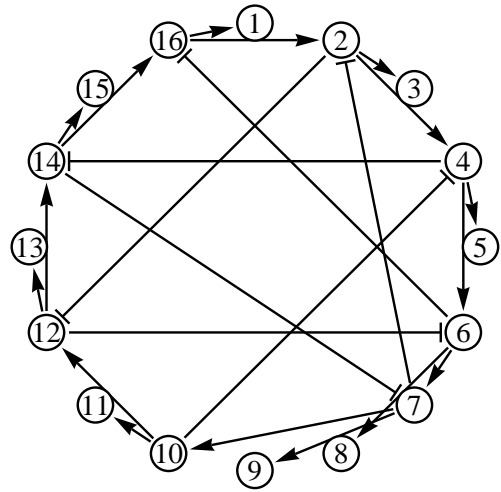}
\caption{\label{fig:circuits_cycle} 
The most common essential network topology when one imposes
a particular cyclic expression pattern. That pattern corresponds
to a group of active genes that shifts clockwise at each step.
One sees very clearly the activating interactions acting
downstream and the inhibitory interactions acting upstream.
(Data for $N=16$; this topology arises for over $15\%$ of
the networks.)}
\end{figure}
%
   
\subsection{Functionality leads to motif selection}
Working with the full description of genotypes 
is cumbersome and difficult, whereas 
focusing just on essential networks provides
a great deal of intuition, in particular for what
features are relevant for functionality.
The price to pay for this simplicity is some
loss of information; for example, two interactions separately
may be non-essential but nevertheless if one removes 
both of them the network's functionality may be lost.

To obtain insights into network structure,
one can search for network motifs; this has become very
popular in recent years, to a large extent through the effort of
U. Alon and collaborators \cite{MiloShen-Orr2002,AlonBook2006} 
(see also the web page www.weizmann.ac.il/mcb/UriAlon/). The fact that 
a complex network can be constructed from small standard sub-elements is
by itself not surprising. For example, this property is at the
root of electronics and is based on the mathematical structure of
logical functions. However, the fact that nature also uses this strategy
is not obvious, and that some
motifs and not others are employed in different network functions is
even less obvious. This presence of motifs
is revealed though from the detailed 
studies of (rather rare, for obvious reasons) biological networks 
reconstructed from data, and it has been partly explained by 
arguments borrowed from communication systems techniques. 
This brings us to 
inquire what happens in a model where the same dynamics is always
at work, and where thousands of networks can be generated for
several network functions: will the same motifs emerge when
the functionality constraints are modified, or will the motifs
change with the functions implemented by the networks.

\begin{figure}
\centering
\includegraphics[width=8cm]{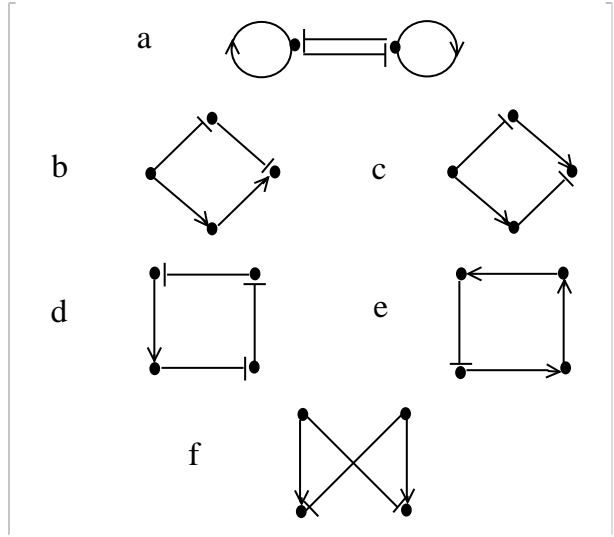}
\caption{\label{fig:motifs_all} 
The most prominent motifs found for our two classes
of functional constraints. 
Case of multi-stability (more than one steady-state): (a) two mutually inhibiting and 
self-activating genes.
Case of expression targets that are cyclic 
in time: (b,c) diamond motif, (d,e) four-node loop, (f) bifan.
 }
\end{figure}

To answer the previous question, we determine the motifs
in our different ensembles. The web page mentioned above offers 
a software for motif search; it
is not quite adapted to our needs, since it does not distinguish between
activators and repressors, and does not accept self-interactions. However,
it was helpful in this work, enabling us to single out the relevant
motif topologies (when a topology is irrelevant, it is also so
when more detailed distinctions are introduced). Furthermore, we used
it to test our own codes for motif extraction.
The results presented here concern the most prominent motifs; others have 
frequencies that are either very small or at least roughly of the 
order of the expectation for a randomized network. We discard motifs with 
leaves (degree-one nodes), which are somewhat trivial. 
We keep only motifs that are not a subgraph of a larger motif with 
the same number of nodes. However, our motifs can partly overlap. The 
randomization used is that proposed by Maslov 
and Sneppen \cite{MaslovSneppen2002}:
the links are interchanged, so that both the in- and out-degrees 
of network nodes
remain unchanged. Our results are summarized in Table~\ref{tab2} 
and the motifs are listed and defined in Fig.~\ref{fig:motifs_all}. 
\begin{center}
\begin{table}
\caption{\label{tab2} Most important motifs.
nFP means ``n fixed points phenotype''. Cycle 
refers to our 8-step cycle.
We typically used 1000 GRNs in our motif
search.
}
\vspace{0.4cm}
\begin{center}
\begin{tabular}{|l|l|l|l|l|} \hline \hline
motif  & 2FP & 3FP & 4FP & cycle\\ \hline
& & &  &\\
motif a: model  & {\bf 0.706(16)} &  {\bf 2.358(39)} & {\bf 2.984(4)}  &  0.000\\
randomized      & 0.002(1)  &  0.002(1) & 0.002(2)  &  0.000\\
\hline
motif b: model  &  0.000    &  0.000    & 0.000     &  {\bf 5.451(41)}\\
randomized      &  0.001(1) & 0.008(3)  & 0.071(9)  &  0.023(5)\\
\hline
motif c: model  &  0.000    & 0.000     & 0.000     &  {\bf 5.170}(40)\\
randomized      &  0.000    & 0.008(3)  & 0.078(9)  &  0.029(1)\\
\hline
motif d: model  &  0.000    & 0.000     & 0.000     &  {\bf 4.533(42)}\\
randomized      &  0.000    & 0.014(4)  & 0.0180(6) &  0.030(7)\\
\hline
motif e: model  &  0.000    & 0.000     & 0.000     &  {\bf 6.676(22)}\\
randomized      &  0.017(4) &  0.102(16) & 0.057(8)  &  0.173(14)\\
\hline
motif f:        &  0.000    & 0.000     & 0.000     &  {\bf 2.296(29)}\\
randomized      &  0.000    & 0.001(1)  & 0.050(6)  &  0.003(2)\\
 & & & &\\
\hline \hline
\end{tabular}
\end{center}
\end{table}
\end{center}
We see right away a very strong dichotomy: the motifs 
are very different for our two classes of functions (imposing
multi-stability vs. cycling). In the case of multi-stability,
one single motif stands out as being extremely important: 
two genes that are mutually inhibitory and which are self-activating.
Clearly such a pair of genes can act as a switch that will then
influence downstream genes according to the expression pattern that
is required for the considered fixed point. When dealing with
more than two target fixed points, additional
such motifs should be necessary. That is indeed what we
saw in Fig.~\ref{fig:circuits_nFP} which displayed
the most represented essential network (ignoring 
permutations of indices) for 2 and 4 imposed fixed points.
Interestingly, the same trend also emerges
for the less frequent essential networks (data not shown).
Roughly, the networks display
a core of central genes that belong to a motif of type ``a'' (using the
nomenclature shown in Fig.~\ref{fig:motifs_all}) and these genes then
influence other genes by a simple downstream effect along the associated
tree-like graph of activating interactions.

Now when we look at the motifs present when imposing cyclic expression
targets, the previous motif is absent and instead we have several
four gene motifs that are strongly over-represented. Motif ``f''
is the bifan in the nomenclature of Alon; the others generally
involve a regulatory loop, and in particular motifs ``d'' and ``e''
are ``frustrated'' in that they have loops with an odd number of
inhibitory interactions. None of these were over-represented
in the networks satisfying multi-stability. Their presence can be understood
by looking at the dominant essential networks such as in 
Fig.~\ref{fig:circuits_cycle}.
Note that these motifs are analogous to the 
ones driving repressilators~\cite{BarkaiLeibler_2000} but
they involve more genes and do not provide oscillatory
behavior on their own, their function requires the
presence of the other genes that are not included in the motif construction.

\section{Discussion and conclusions}

The central question tackled by the present work is whether the
emergence of motifs in gene regulatory networks can be 
due to functional constraints. Given the uncertainties in 
how real genetic networks function, we have taken a modeling route
and have addressed this question \emph{in silico}. Our model incorporates
known molecular mechanisms for the description of genetic interactions,
and in fact the only parameters in our model come from
parametrizing the affinities of transcription factors to their
binding sites. Furthermore, in contrast to most other gene 
regulatory network modelings, the associated interactions  
are never completely absent; they can be important or unimportant
for the functionality of the network, a notion we 
characterized by the \emph{essentiality} of interactions.
Finally, the expression level of each
gene follows dynamics allowing for continuous values; this 
additional complexity compared 
to using digital ``on-off'' expression levels forces one to 
consider functionality as a soft constraint, imposing 
expression levels to be ``sufficiently'' close to target
patterns. Network functionality is then quantified via a 
fitness measure. Such a framework provides a close parallel with
thermodynamic ensembles; all questions are then necessarily
posed in a probabilistic framework where each network
arises with a probability proportional to its fitness.
In practice, we explore the corresponding ensemble of genetic 
networks numerically, using Monte Carlo Markov Chain.

Two types of gene network functionalities have been studied.
The first is motivated by the different cell types in 
multi-cellular organisms and is implemented by constraining
the genes in the networks to have steady-state expression levels 
close to given target levels; in effect, the transcriptional
dynamics of the networks must allow for multi-stability, that is
multiple fixed points of the expression dynamics.
The second type of functional constraint considered is motivated
by previous work on the cell cycle; we implemented such a constraint by
forcing the networks to have their expression levels 
follow a given cyclic pattern in time. Thus instead of fixed
points, in this case we ask for a periodic behavior of the dynamics.
In both cases, we found characteristic features shared with
other models of living systems~\cite{WagnerBook2005} as follows. (1) 
The constraints imposed are extremely stringent as can be seen from
the fact that in practice they are never satisfied by 
randomly generated networks. (2) Although the fraction of networks of 
interest is tiny, the number of networks satisfying the constraints is 
astronomical as revealed by our Monte Carlo Markov Chain sampling.

Of interest is the structure of these presumably atypical
networks. Particular architectures are known to arise
when performing genetic network design via optimization 
algorithms~\cite{FrancoisHakim2004,FrancoisHakim2007,RodrigoCarrera2007}.
Is this property a bias of these algorithms or does it reflect
an underlying constraint imposed by network function? It is difficult
to tackle this question head-on except in very small systems; there
one can explore all possible values for the
model's parameters~\cite{GuantesPoyatos_2008} and see
the functional consequences. Since motifs can involve three or
even more genes inside a larger network, a different approach is
necessary for moderate and large networks. The most adapted tool 
is based on Monte Carlo Markov chains and so we have
applied this approach to our systems with up to 16 genes. MCMC
then allows us to sample the space of \emph{functional} networks
in spite of the fact that it represents only a tiny fraction
of the space of \emph{all} networks.

Given a gene regulatory network
produced by the Monte Carlo algorithm, we first extracted 
the essential interactions to obtain what we called
the \emph{essential networks}. This representation gets rid of 
irrelevant interactions that are too small to influence
much the functionality. Interestingly, these essential networks
are sparse and make use of inhibitory interactions
parsimoniously. We then determined the motifs appearing in these
essential networks, where a motif is an oriented sub-graph that is overly
frequent when comparing with a randomization test preserving
each node's degree. In the case of networks satisfying
the multi-stability constraints, we found one very dominant
motif of two genes acting as a switch: each gene represses
the other while activating itself. Furthermore, this motif
arose once when imposing two fixed points to the dynamics,
twice when imposing three fixed points to the dynamics etc.
This pattern makes good sense from a ``design'' perspective: 
the choice to go to one fixed point rather than to
another can be implemented
most simply by using switches that operate in this logical fashion.
Moving on now to the ensemble of networks that implemented
expression patterns that were cyclic in time, we found here that
the dominant motifs involved 4 genes as shown in Fig.~\ref{fig:motifs_all}.
One of these corresponds to the bifan in Alon's nomenclature, 
but four other motifs were also found and in fact were even more
often present. All of these motifs involve at least one inhibitory
interaction; this is appropriate for our imposed cycle as the newly
turned on genes must at some point turn off the other genes they
are replacing. Interestingly, the motifs we find in one ensemble
are not present in the other. This shows that functionality
is a major determinant of the content in motifs, at least 
within our simplified framework.
Some importance of functionality could have been expected 
\emph{a priori}, but the size of the effect is striking. We hope
this result will encourage the search for functional biases
between experimental motifs, in particular through comparative studies.


\section*{Acknowledgments}
We thank V. Fromion, L. Giorgetti and V. Hakim for helpful comments.
This work was supported
by the Polish Ministry of Science
Grant No.~N~N202~229137 (2009-2012). 
The project operated within the Foundation for Polish Science 
International Ph.D. Projects Programme co-financed by the European 
Regional Development Fund, agreement no. MPD/2009/6.
The LPT, LPTMS, and UMR de G\'en\'etique V\'eg\'etale
are Unit\'e de Recherche de
l'Universit\'e Paris-Sud associ\'ees au CNRS. Marcin Zagorski is grateful to LPT for hospitality.



\addcontentsline{toc}{chapter}{\protect\bibname}
\begin{thebibliography}{45}%
\makeatletter
\providecommand \@ifxundefined [1]{%
 \@ifx{#1\undefined}
}%
\providecommand \@ifnum [1]{%
 \ifnum #1\expandafter \@firstoftwo
 \else \expandafter \@secondoftwo
 \fi
}%
\providecommand \@ifx [1]{%
 \ifx #1\expandafter \@firstoftwo
 \else \expandafter \@secondoftwo
 \fi
}%
\providecommand \natexlab [1]{#1}%
\providecommand \enquote  [1]{``#1''}%
\providecommand \bibnamefont  [1]{#1}%
\providecommand \bibfnamefont [1]{#1}%
\providecommand \citenamefont [1]{#1}%
\providecommand \href@noop [0]{\@secondoftwo}%
\providecommand \href [0]{\begingroup \@sanitize@url \@href}%
\providecommand \@href[1]{\@@startlink{#1}\@@href}%
\providecommand \@@href[1]{\endgroup#1\@@endlink}%
\providecommand \@sanitize@url [0]{\catcode `\\12\catcode `\$12\catcode
  `\&12\catcode `\#12\catcode `\^12\catcode `\_12\catcode `\%12\relax}%
\providecommand \@@startlink[1]{}%
\providecommand \@@endlink[0]{}%
\providecommand \url  [0]{\begingroup\@sanitize@url \@url }%
\providecommand \@url [1]{\endgroup\@href {#1}{\urlprefix }}%
\providecommand \urlprefix  [0]{URL }%
\providecommand \Eprint [0]{\href }%
\@ifxundefined \urlstyle {%
  \providecommand \doi  [0]{\begingroup \@sanitize@url \@doi}%
  \providecommand \@doi [1]{\endgroup \@@startlink {\doibase
  #1}doi:\discretionary {}{}{}#1\@@endlink }%
}{%
  \providecommand \doi  [0]{doi:\discretionary{}{}{}\begingroup
  \urlstyle{rm}\Url }%
}%
\providecommand \doibase [0]{http://dx.doi.org/}%
\providecommand \Doi [0]{\begingroup \@sanitize@url \@Doi }%
\providecommand \@Doi  [1]{\endgroup\@@startlink{\doibase#1}\@@Doi}%
\providecommand \@@Doi [1]{#1\@@endlink}%
\providecommand \selectlanguage [0]{\@gobble}%
\providecommand \bibinfo  [0]{\@secondoftwo}%
\providecommand \bibfield  [0]{\@secondoftwo}%
\providecommand \translation [1]{[#1]}%
\providecommand \BibitemOpen [0]{}%
\providecommand \bibitemStop [0]{}%
\providecommand \bibitemNoStop [0]{.\EOS\space}%
\providecommand \EOS [0]{\spacefactor3000\relax}%
\providecommand \BibitemShut  [1]{\csname bibitem#1\endcsname}%
\bibitem [{\citenamefont {Elowitz}\ and\ \citenamefont
  {Leibler}(2000)}]{Elowitz_Leibler_2000}%
  \BibitemOpen
  \bibfield  {author} {\bibinfo {author} {\bibfnamefont {M.}~\bibnamefont
  {Elowitz}}\ and\ \bibinfo {author} {\bibfnamefont {S.}~\bibnamefont
  {Leibler}},\ }\href@noop {} {\bibfield  {journal} {\bibinfo  {journal}
  {Nature},\ }\textbf {\bibinfo {volume} {403}},\ \bibinfo {pages} {335}
  (\bibinfo {year} {2000})}\BibitemShut {NoStop}%
\bibitem [{\citenamefont {Gardner}\ \emph {et~al.}(2000)\citenamefont
  {Gardner}, \citenamefont {Cantor},\ and\ \citenamefont
  {Collins}}]{Gardner_Cantor_2000}%
  \BibitemOpen
  \bibfield  {author} {\bibinfo {author} {\bibfnamefont {T.}~\bibnamefont
  {Gardner}}, \bibinfo {author} {\bibfnamefont {C.}~\bibnamefont {Cantor}}, \
  and\ \bibinfo {author} {\bibfnamefont {J.}~\bibnamefont {Collins}},\
  }\href@noop {} {\bibfield  {journal} {\bibinfo  {journal} {Nature},\ }\textbf
  {\bibinfo {volume} {403}},\ \bibinfo {pages} {339} (\bibinfo {year}
  {2000})}\BibitemShut {NoStop}%
\bibitem [{\citenamefont {Herrgard}\ \emph {et~al.}(2004)\citenamefont
  {Herrgard}, \citenamefont {Covert},\ and\ \citenamefont
  {Palsson}}]{HerrgardCovert2004}%
  \BibitemOpen
  \bibfield  {author} {\bibinfo {author} {\bibfnamefont {M.}~\bibnamefont
  {Herrgard}}, \bibinfo {author} {\bibfnamefont {M.}~\bibnamefont {Covert}}, \
  and\ \bibinfo {author} {\bibfnamefont {B.}~\bibnamefont {Palsson}},\
  }\href@noop {} {\bibfield  {journal} {\bibinfo  {journal} {Current Opinion in
  Biotechnology},\ }\textbf {\bibinfo {volume} {15(1)}},\ \bibinfo {pages} {70}
  (\bibinfo {year} {2004})}\BibitemShut {NoStop}%
\bibitem [{\citenamefont {Salgado}\ \emph {et~al.}(2006)\citenamefont
  {Salgado}, \citenamefont {Gama-Castro}, \citenamefont {Peralta-Gil},
  \citenamefont {D\'iaz-Peredo}, \citenamefont {S\'anchez-Solano},
  \citenamefont {Santos-Zavaleta}, \citenamefont {Mart\'inez-Flores},
  \citenamefont {Jim\'enez-Jacinto}, \citenamefont {Bonavides-Mart\'inez},
  \citenamefont {Segura-Salazar}, \citenamefont {Mart\'inez-Antonio},\ and\
  \citenamefont {Collado-Vides}}]{SalgadoGama2006}%
  \BibitemOpen
  \bibfield  {author} {\bibinfo {author} {\bibfnamefont {H.}~\bibnamefont
  {Salgado}}, \bibinfo {author} {\bibfnamefont {S.}~\bibnamefont
  {Gama-Castro}}, \bibinfo {author} {\bibfnamefont {M.}~\bibnamefont
  {Peralta-Gil}}, \bibinfo {author} {\bibfnamefont {E.}~\bibnamefont
  {D\'iaz-Peredo}}, \bibinfo {author} {\bibfnamefont {F.}~\bibnamefont
  {S\'anchez-Solano}}, \bibinfo {author} {\bibfnamefont {A.}~\bibnamefont
  {Santos-Zavaleta}}, \bibinfo {author} {\bibfnamefont {I.}~\bibnamefont
  {Mart\'inez-Flores}}, \bibinfo {author} {\bibfnamefont {V.}~\bibnamefont
  {Jim\'enez-Jacinto}}, \bibinfo {author} {\bibfnamefont {C.}~\bibnamefont
  {Bonavides-Mart\'inez}}, \bibinfo {author} {\bibfnamefont {J.}~\bibnamefont
  {Segura-Salazar}}, \bibinfo {author} {\bibfnamefont {A.}~\bibnamefont
  {Mart\'inez-Antonio}}, \ and\ \bibinfo {author} {\bibfnamefont
  {J.}~\bibnamefont {Collado-Vides}},\ }\href@noop {} {\bibfield  {journal}
  {\bibinfo  {journal} {Nucleic Acids Research},\ }\textbf {\bibinfo {volume}
  {34}},\ \bibinfo {pages} {D394} (\bibinfo {year} {2006})}\BibitemShut
  {NoStop}%
\bibitem [{\citenamefont {Hu}\ \emph {et~al.}(2007)\citenamefont {Hu},
  \citenamefont {Killion},\ and\ \citenamefont {Iyer}}]{Hu_Killion_2007}%
  \BibitemOpen
  \bibfield  {author} {\bibinfo {author} {\bibfnamefont {Z.}~\bibnamefont
  {Hu}}, \bibinfo {author} {\bibfnamefont {P.}~\bibnamefont {Killion}}, \ and\
  \bibinfo {author} {\bibfnamefont {V.}~\bibnamefont {Iyer}},\ }\href@noop {}
  {\bibfield  {journal} {\bibinfo  {journal} {Nature Genetics},\ }\textbf
  {\bibinfo {volume} {39}},\ \bibinfo {pages} {683} (\bibinfo {year}
  {2007})}\BibitemShut {NoStop}%
\bibitem [{\citenamefont {Shen-Orr}\ \emph {et~al.}(2002)\citenamefont
  {Shen-Orr}, \citenamefont {Milo}, \citenamefont {Mangan},\ and\ \citenamefont
  {Alon}}]{Shen-Orr_Milo_2002}%
  \BibitemOpen
  \bibfield  {author} {\bibinfo {author} {\bibfnamefont {S.}~\bibnamefont
  {Shen-Orr}}, \bibinfo {author} {\bibfnamefont {R.}~\bibnamefont {Milo}},
  \bibinfo {author} {\bibfnamefont {S.}~\bibnamefont {Mangan}}, \ and\ \bibinfo
  {author} {\bibfnamefont {U.}~\bibnamefont {Alon}},\ }\href@noop {} {\bibfield
   {journal} {\bibinfo  {journal} {Nature Genetics},\ }\textbf {\bibinfo
  {volume} {31}},\ \bibinfo {pages} {64} (\bibinfo {year} {2002})}\BibitemShut
  {NoStop}%
\bibitem [{\citenamefont {Ma}\ \emph {et~al.}(2004)\citenamefont {Ma},
  \citenamefont {Kumar}, \citenamefont {Ditges}, \citenamefont {Gunzer},
  \citenamefont {Buer},\ and\ \citenamefont {Zeng}}]{Ma_Kumar_2004}%
  \BibitemOpen
  \bibfield  {author} {\bibinfo {author} {\bibfnamefont {H.}~\bibnamefont
  {Ma}}, \bibinfo {author} {\bibfnamefont {B.}~\bibnamefont {Kumar}}, \bibinfo
  {author} {\bibfnamefont {U.}~\bibnamefont {Ditges}}, \bibinfo {author}
  {\bibfnamefont {F.}~\bibnamefont {Gunzer}}, \bibinfo {author} {\bibfnamefont
  {J.}~\bibnamefont {Buer}}, \ and\ \bibinfo {author} {\bibfnamefont
  {A.}~\bibnamefont {Zeng}},\ }\href@noop {} {\bibfield  {journal} {\bibinfo
  {journal} {Nucl. Acids Res.},\ }\textbf {\bibinfo {volume} {32(22)}},\
  \bibinfo {pages} {6643} (\bibinfo {year} {2004})}\BibitemShut {NoStop}%
\bibitem [{\citenamefont {Lee}\ \emph {et~al.}(2002)\citenamefont {Lee},
  \citenamefont {Rinaldi}, \citenamefont {Robert}, \citenamefont {Odom},
  \citenamefont {Bar-Joseph}, \citenamefont {Gerber}, \citenamefont {Hannett},
  \citenamefont {Harbison}, \citenamefont {Thompson}, \citenamefont {Simon},
  \citenamefont {Zeitlinger}, \citenamefont {Jennings}, \citenamefont {Murray},
  \citenamefont {Gordon}, \citenamefont {Ren}, \citenamefont {Wyrick},
  \citenamefont {Tagne}, \citenamefont {Volkert}, \citenamefont {Fraenkel},
  \citenamefont {Gifford},\ and\ \citenamefont {Young}}]{Lee_Rinaldi_2002}%
  \BibitemOpen
  \bibfield  {author} {\bibinfo {author} {\bibfnamefont {T.}~\bibnamefont
  {Lee}}, \bibinfo {author} {\bibfnamefont {N.}~\bibnamefont {Rinaldi}},
  \bibinfo {author} {\bibfnamefont {F.}~\bibnamefont {Robert}}, \bibinfo
  {author} {\bibfnamefont {D.}~\bibnamefont {Odom}}, \bibinfo {author}
  {\bibfnamefont {Z.}~\bibnamefont {Bar-Joseph}}, \bibinfo {author}
  {\bibfnamefont {G.}~\bibnamefont {Gerber}}, \bibinfo {author} {\bibfnamefont
  {N.}~\bibnamefont {Hannett}}, \bibinfo {author} {\bibfnamefont
  {C.}~\bibnamefont {Harbison}}, \bibinfo {author} {\bibfnamefont
  {C.}~\bibnamefont {Thompson}}, \bibinfo {author} {\bibfnamefont
  {I.}~\bibnamefont {Simon}}, \bibinfo {author} {\bibfnamefont
  {J.}~\bibnamefont {Zeitlinger}}, \bibinfo {author} {\bibfnamefont
  {E.}~\bibnamefont {Jennings}}, \bibinfo {author} {\bibfnamefont
  {H.}~\bibnamefont {Murray}}, \bibinfo {author} {\bibfnamefont
  {D.}~\bibnamefont {Gordon}}, \bibinfo {author} {\bibfnamefont
  {B.}~\bibnamefont {Ren}}, \bibinfo {author} {\bibfnamefont {J.}~\bibnamefont
  {Wyrick}}, \bibinfo {author} {\bibfnamefont {J.}~\bibnamefont {Tagne}},
  \bibinfo {author} {\bibfnamefont {T.}~\bibnamefont {Volkert}}, \bibinfo
  {author} {\bibfnamefont {E.}~\bibnamefont {Fraenkel}}, \bibinfo {author}
  {\bibfnamefont {D.}~\bibnamefont {Gifford}}, \ and\ \bibinfo {author}
  {\bibfnamefont {R.}~\bibnamefont {Young}},\ }\href@noop {} {\bibfield
  {journal} {\bibinfo  {journal} {Science},\ }\textbf {\bibinfo {volume}
  {298}},\ \bibinfo {pages} {799} (\bibinfo {year} {2002})}\BibitemShut
  {NoStop}%
\bibitem [{\citenamefont {Zhu}\ \emph {et~al.}(2008)\citenamefont {Zhu},
  \citenamefont {Zhang}, \citenamefont {Smith}, \citenamefont {Drees},
  \citenamefont {Brem}, \citenamefont {Kruglyak}, \citenamefont {Bumgarner},\
  and\ \citenamefont {Schadt}}]{ZhuZhang2008}%
  \BibitemOpen
  \bibfield  {author} {\bibinfo {author} {\bibfnamefont {J.}~\bibnamefont
  {Zhu}}, \bibinfo {author} {\bibfnamefont {B.}~\bibnamefont {Zhang}}, \bibinfo
  {author} {\bibfnamefont {E.}~\bibnamefont {Smith}}, \bibinfo {author}
  {\bibfnamefont {B.}~\bibnamefont {Drees}}, \bibinfo {author} {\bibfnamefont
  {R.}~\bibnamefont {Brem}}, \bibinfo {author} {\bibfnamefont {L.}~\bibnamefont
  {Kruglyak}}, \bibinfo {author} {\bibfnamefont {R.}~\bibnamefont {Bumgarner}},
  \ and\ \bibinfo {author} {\bibfnamefont {E.}~\bibnamefont {Schadt}},\
  }\href@noop {} {\bibfield  {journal} {\bibinfo  {journal} {Nature Genetics},\
  }\textbf {\bibinfo {volume} {40}},\ \bibinfo {pages} {854} (\bibinfo {year}
  {2008})}\BibitemShut {NoStop}%
\bibitem [{\citenamefont {Mangan}\ and\ \citenamefont
  {Alon}(2003)}]{Mangan_Alon_2003}%
  \BibitemOpen
  \bibfield  {author} {\bibinfo {author} {\bibfnamefont {S.}~\bibnamefont
  {Mangan}}\ and\ \bibinfo {author} {\bibfnamefont {U.}~\bibnamefont {Alon}},\
  }\href@noop {} {\bibfield  {journal} {\bibinfo  {journal} {Proc. Natl. Acad.
  Sci.},\ }\textbf {\bibinfo {volume} {100(21)}},\ \bibinfo {pages} {11980}
  (\bibinfo {year} {2003})}\BibitemShut {NoStop}%
\bibitem [{\citenamefont {Camas}\ and\ \citenamefont
  {Poyatos}(2008)}]{Camas_Poyatos_2008}%
  \BibitemOpen
  \bibfield  {author} {\bibinfo {author} {\bibfnamefont {F.}~\bibnamefont
  {Camas}}\ and\ \bibinfo {author} {\bibfnamefont {J.}~\bibnamefont
  {Poyatos}},\ }\href@noop {} {\bibfield  {journal} {\bibinfo  {journal} {PLoS
  One},\ }\textbf {\bibinfo {volume} {3(11)}},\ \bibinfo {pages} {e3657}
  (\bibinfo {year} {2008})}\BibitemShut {NoStop}%
\bibitem [{\citenamefont {Fran\c{c}ois}\ and\ \citenamefont
  {Hakim}(2004)}]{FrancoisHakim2004}%
  \BibitemOpen
  \bibfield  {author} {\bibinfo {author} {\bibfnamefont {P.}~\bibnamefont
  {Fran\c{c}ois}}\ and\ \bibinfo {author} {\bibfnamefont {V.}~\bibnamefont
  {Hakim}},\ }\href@noop {} {\bibfield  {journal} {\bibinfo  {journal} {Proc.
  Natl. Acad. Sci.},\ }\textbf {\bibinfo {volume} {101}},\ \bibinfo {pages}
  {580} (\bibinfo {year} {2004})}\BibitemShut {NoStop}%
\bibitem [{\citenamefont {Fran\c{c}ois}\ \emph {et~al.}(2007)\citenamefont
  {Fran\c{c}ois}, \citenamefont {Hakim},\ and\ \citenamefont
  {Siggia}}]{FrancoisHakim2007}%
  \BibitemOpen
  \bibfield  {author} {\bibinfo {author} {\bibfnamefont {P.}~\bibnamefont
  {Fran\c{c}ois}}, \bibinfo {author} {\bibfnamefont {V.}~\bibnamefont {Hakim}},
  \ and\ \bibinfo {author} {\bibfnamefont {E.}~\bibnamefont {Siggia}},\
  }\href@noop {} {\bibfield  {journal} {\bibinfo  {journal} {Mol. Syst.
  Biol.},\ }\textbf {\bibinfo {volume} {3}},\ \bibinfo {pages} {154} (\bibinfo
  {year} {2007})}\BibitemShut {NoStop}%
\bibitem [{\citenamefont {Rodrigo}\ \emph {et~al.}(2007)\citenamefont
  {Rodrigo}, \citenamefont {Carrera},\ and\ \citenamefont
  {Jaramillo}}]{RodrigoCarrera2007}%
  \BibitemOpen
  \bibfield  {author} {\bibinfo {author} {\bibfnamefont {G.}~\bibnamefont
  {Rodrigo}}, \bibinfo {author} {\bibfnamefont {J.}~\bibnamefont {Carrera}}, \
  and\ \bibinfo {author} {\bibfnamefont {A.}~\bibnamefont {Jaramillo}},\
  }\href@noop {} {\bibfield  {journal} {\bibinfo  {journal} {Bioinformatics},\
  }\textbf {\bibinfo {volume} {23}},\ \bibinfo {pages} {1857} (\bibinfo {year}
  {2007})}\BibitemShut {NoStop}%
\bibitem [{\citenamefont {Alon}(2007)}]{AlonBook2006}%
  \BibitemOpen
  \bibfield  {author} {\bibinfo {author} {\bibfnamefont {U.}~\bibnamefont
  {Alon}},\ }\href@noop {} {\emph {\bibinfo {title} {An Introduction to Systems
  Biology: Design Principles of Biological Circuits}}}\ (\bibinfo  {publisher}
  {Chapman and Hall},\ \bibinfo {address} {Boca Raton, FL},\ \bibinfo {year}
  {2007})\BibitemShut {NoStop}%
\bibitem [{\citenamefont {Burda}\ \emph {et~al.}(2010)\citenamefont {Burda},
  \citenamefont {Krzywicki}, \citenamefont {Martin},\ and\ \citenamefont
  {Zagorski}}]{BKMZ_2010}%
  \BibitemOpen
  \bibfield  {author} {\bibinfo {author} {\bibfnamefont {Z.}~\bibnamefont
  {Burda}}, \bibinfo {author} {\bibfnamefont {A.}~\bibnamefont {Krzywicki}},
  \bibinfo {author} {\bibfnamefont {O.}~\bibnamefont {Martin}}, \ and\ \bibinfo
  {author} {\bibfnamefont {M.}~\bibnamefont {Zagorski}},\ }\href@noop {}
  {\bibfield  {journal} {\bibinfo  {journal} {Physical Review E},\ }\textbf
  {\bibinfo {volume} {82}},\ \bibinfo {pages} {011908} (\bibinfo {year}
  {2010})}\BibitemShut {NoStop}%
\bibitem [{\citenamefont {von Hippel}\ and\ \citenamefont
  {Berg}(1986)}]{HippelBerg1986}%
  \BibitemOpen
  \bibfield  {author} {\bibinfo {author} {\bibfnamefont {P.}~\bibnamefont {von
  Hippel}}\ and\ \bibinfo {author} {\bibfnamefont {O.}~\bibnamefont {Berg}},\
  }\href@noop {} {\bibfield  {journal} {\bibinfo  {journal} {Proc. Natl. Acad.
  Sci.},\ }\textbf {\bibinfo {volume} {83}},\ \bibinfo {pages} {1608} (\bibinfo
  {year} {1986})}\BibitemShut {NoStop}%
\bibitem [{\citenamefont {Berg}\ and\ \citenamefont {von
  Hippel}(1987)}]{BergHippel1987}%
  \BibitemOpen
  \bibfield  {author} {\bibinfo {author} {\bibfnamefont {O.}~\bibnamefont
  {Berg}}\ and\ \bibinfo {author} {\bibfnamefont {P.}~\bibnamefont {von
  Hippel}},\ }\href@noop {} {\bibfield  {journal} {\bibinfo  {journal} {J. Mol.
  Biol.},\ }\textbf {\bibinfo {volume} {193}},\ \bibinfo {pages} {723}
  (\bibinfo {year} {1987})}\BibitemShut {NoStop}%
\bibitem [{\citenamefont {Gerland}\ \emph {et~al.}(2002)\citenamefont
  {Gerland}, \citenamefont {Moroz},\ and\ \citenamefont
  {Hwa}}]{GerlandMoroz2002}%
  \BibitemOpen
  \bibfield  {author} {\bibinfo {author} {\bibfnamefont {U.}~\bibnamefont
  {Gerland}}, \bibinfo {author} {\bibfnamefont {J.}~\bibnamefont {Moroz}}, \
  and\ \bibinfo {author} {\bibfnamefont {T.}~\bibnamefont {Hwa}},\ }\href@noop
  {} {\bibfield  {journal} {\bibinfo  {journal} {Proc. Natl. Acad. Sci.},\
  }\textbf {\bibinfo {volume} {99}},\ \bibinfo {pages} {12015} (\bibinfo {year}
  {2002})}\BibitemShut {NoStop}%
\bibitem [{\citenamefont {Sarai}\ and\ \citenamefont
  {Takeda}(1989)}]{SaraiTakeda1989}%
  \BibitemOpen
  \bibfield  {author} {\bibinfo {author} {\bibfnamefont {A.}~\bibnamefont
  {Sarai}}\ and\ \bibinfo {author} {\bibfnamefont {Y.}~\bibnamefont {Takeda}},\
  }\href@noop {} {\bibfield  {journal} {\bibinfo  {journal} {Proc. Natl. Acad.
  Sci.},\ }\textbf {\bibinfo {volume} {86}},\ \bibinfo {pages} {6513} (\bibinfo
  {year} {1989})}\BibitemShut {NoStop}%
\bibitem [{\citenamefont {Stormo}\ and\ \citenamefont
  {Fields}(1998)}]{StormoFields1998}%
  \BibitemOpen
  \bibfield  {author} {\bibinfo {author} {\bibfnamefont {G.}~\bibnamefont
  {Stormo}}\ and\ \bibinfo {author} {\bibfnamefont {D.}~\bibnamefont
  {Fields}},\ }\href@noop {} {\bibfield  {journal} {\bibinfo  {journal} {Trends
  in Biochem. Sci.},\ }\textbf {\bibinfo {volume} {23}},\ \bibinfo {pages}
  {109} (\bibinfo {year} {1998})}\BibitemShut {NoStop}%
\bibitem [{\citenamefont {Bulyk}\ \emph {et~al.}(2002)\citenamefont {Bulyk},
  \citenamefont {Johnson},\ and\ \citenamefont {Church}}]{BulykJohnson2002}%
  \BibitemOpen
  \bibfield  {author} {\bibinfo {author} {\bibfnamefont {M.}~\bibnamefont
  {Bulyk}}, \bibinfo {author} {\bibfnamefont {P.}~\bibnamefont {Johnson}}, \
  and\ \bibinfo {author} {\bibfnamefont {G.}~\bibnamefont {Church}},\
  }\href@noop {} {\bibfield  {journal} {\bibinfo  {journal} {Nucl. Acids
  Res.},\ }\textbf {\bibinfo {volume} {20}},\ \bibinfo {pages} {1255} (\bibinfo
  {year} {2002})}\BibitemShut {NoStop}%
\bibitem [{\citenamefont {Granek}\ and\ \citenamefont
  {Clark}(2005)}]{Granek_Clarke_2005}%
  \BibitemOpen
  \bibfield  {author} {\bibinfo {author} {\bibfnamefont {J.}~\bibnamefont
  {Granek}}\ and\ \bibinfo {author} {\bibfnamefont {N.}~\bibnamefont {Clark}},\
  }\href@noop {} {\bibfield  {journal} {\bibinfo  {journal} {Genome Biology},\
  }\textbf {\bibinfo {volume} {6}},\ \bibinfo {pages} {R87} (\bibinfo {year}
  {2005})}\BibitemShut {NoStop}%
\bibitem [{\citenamefont {Golding}\ \emph {et~al.}(2005)\citenamefont
  {Golding}, \citenamefont {Paulsson}, \citenamefont {Zawilski},\ and\
  \citenamefont {Cox}}]{GoldingPaulsson2005}%
  \BibitemOpen
  \bibfield  {author} {\bibinfo {author} {\bibfnamefont {I.}~\bibnamefont
  {Golding}}, \bibinfo {author} {\bibfnamefont {J.}~\bibnamefont {Paulsson}},
  \bibinfo {author} {\bibfnamefont {S.}~\bibnamefont {Zawilski}}, \ and\
  \bibinfo {author} {\bibfnamefont {E.}~\bibnamefont {Cox}},\ }\href@noop {}
  {\bibfield  {journal} {\bibinfo  {journal} {Cell},\ }\textbf {\bibinfo
  {volume} {123}},\ \bibinfo {pages} {1025} (\bibinfo {year}
  {2005})}\BibitemShut {NoStop}%
\bibitem [{\citenamefont {Becskei}\ \emph {et~al.}(2004)\citenamefont
  {Becskei}, \citenamefont {Kaufmann},\ and\ \citenamefont {van
  Oudenaarden}}]{BecskeiKaufmann2005}%
  \BibitemOpen
  \bibfield  {author} {\bibinfo {author} {\bibfnamefont {A.}~\bibnamefont
  {Becskei}}, \bibinfo {author} {\bibfnamefont {B.}~\bibnamefont {Kaufmann}}, \
  and\ \bibinfo {author} {\bibfnamefont {A.}~\bibnamefont {van Oudenaarden}},\
  }\href@noop {} {\bibfield  {journal} {\bibinfo  {journal} {Nat. Rev.
  Genet.},\ }\textbf {\bibinfo {volume} {5}},\ \bibinfo {pages} {101} (\bibinfo
  {year} {2004})}\BibitemShut {NoStop}%
\bibitem [{\citenamefont {Elf}\ \emph {et~al.}(2007)\citenamefont {Elf},
  \citenamefont {Li},\ and\ \citenamefont {Xi}}]{ElfLi2007}%
  \BibitemOpen
  \bibfield  {author} {\bibinfo {author} {\bibfnamefont {J.}~\bibnamefont
  {Elf}}, \bibinfo {author} {\bibfnamefont {G.-W.}\ \bibnamefont {Li}}, \ and\
  \bibinfo {author} {\bibfnamefont {X.}~\bibnamefont {Xi}},\ }\href@noop {}
  {\bibfield  {journal} {\bibinfo  {journal} {Science},\ }\textbf {\bibinfo
  {volume} {316}},\ \bibinfo {pages} {1191} (\bibinfo {year}
  {2007})}\BibitemShut {NoStop}%
\bibitem [{\citenamefont {Kauffman}(1993)}]{KauffmanBook1993}%
  \BibitemOpen
  \bibfield  {author} {\bibinfo {author} {\bibfnamefont {S.}~\bibnamefont
  {Kauffman}},\ }\href@noop {} {\emph {\bibinfo {title} {Origins of Order:
  Self-Organization and Selection in Evolution}}}\ (\bibinfo  {publisher}
  {Oxford University Press},\ \bibinfo {address} {Oxford},\ \bibinfo {year}
  {1993})\BibitemShut {NoStop}%
\bibitem [{\citenamefont {Wagner}(1996)}]{Wagner1996}%
  \BibitemOpen
  \bibfield  {author} {\bibinfo {author} {\bibfnamefont {A.}~\bibnamefont
  {Wagner}},\ }\href@noop {} {\bibfield  {journal} {\bibinfo  {journal}
  {Evolution},\ }\textbf {\bibinfo {volume} {59}},\ \bibinfo {pages} {1008}
  (\bibinfo {year} {1996})}\BibitemShut {NoStop}%
\bibitem [{\citenamefont {Bornholdt}\ and\ \citenamefont
  {Rohlf}(2000)}]{BornholdtRohlf2000}%
  \BibitemOpen
  \bibfield  {author} {\bibinfo {author} {\bibfnamefont {S.}~\bibnamefont
  {Bornholdt}}\ and\ \bibinfo {author} {\bibfnamefont {T.}~\bibnamefont
  {Rohlf}},\ }\href@noop {} {\bibfield  {journal} {\bibinfo  {journal} {Phys.
  Rev. Lett.},\ }\textbf {\bibinfo {volume} {84}},\ \bibinfo {pages} {6114}
  (\bibinfo {year} {2000})}\BibitemShut {NoStop}%
\bibitem [{\citenamefont {Li}\ \emph {et~al.}(2004)\citenamefont {Li},
  \citenamefont {Long}, \citenamefont {Lu}, \citenamefont {Ouyang},\ and\
  \citenamefont {Tang}}]{LiLong2004}%
  \BibitemOpen
  \bibfield  {author} {\bibinfo {author} {\bibfnamefont {F.}~\bibnamefont
  {Li}}, \bibinfo {author} {\bibfnamefont {T.}~\bibnamefont {Long}}, \bibinfo
  {author} {\bibfnamefont {Y.}~\bibnamefont {Lu}}, \bibinfo {author}
  {\bibfnamefont {Q.}~\bibnamefont {Ouyang}}, \ and\ \bibinfo {author}
  {\bibfnamefont {C.}~\bibnamefont {Tang}},\ }\href@noop {} {\bibfield
  {journal} {\bibinfo  {journal} {Proc. Natl. Acad. Sci.},\ }\textbf {\bibinfo
  {volume} {101}},\ \bibinfo {pages} {4781} (\bibinfo {year}
  {2004})}\BibitemShut {NoStop}%
\bibitem [{\citenamefont {Azevedo}\ \emph {et~al.}(2006)\citenamefont
  {Azevedo}, \citenamefont {Lohaus}, \citenamefont {Srinivasan}, \citenamefont
  {Dang},\ and\ \citenamefont {Burch}}]{AzevedoLohaus2006}%
  \BibitemOpen
  \bibfield  {author} {\bibinfo {author} {\bibfnamefont {R.}~\bibnamefont
  {Azevedo}}, \bibinfo {author} {\bibfnamefont {R.}~\bibnamefont {Lohaus}},
  \bibinfo {author} {\bibfnamefont {S.}~\bibnamefont {Srinivasan}}, \bibinfo
  {author} {\bibfnamefont {K.}~\bibnamefont {Dang}}, \ and\ \bibinfo {author}
  {\bibfnamefont {C.}~\bibnamefont {Burch}},\ }\href@noop {} {\bibfield
  {journal} {\bibinfo  {journal} {Nature},\ }\textbf {\bibinfo {volume}
  {440}},\ \bibinfo {pages} {87} (\bibinfo {year} {2006})}\BibitemShut
  {NoStop}%
\bibitem [{\citenamefont {Segal}\ and\ \citenamefont
  {Widom}(2002)}]{Segal_Widom_2009}%
  \BibitemOpen
  \bibfield  {author} {\bibinfo {author} {\bibfnamefont {E.}~\bibnamefont
  {Segal}}\ and\ \bibinfo {author} {\bibfnamefont {J.}~\bibnamefont {Widom}},\
  }\href@noop {} {\bibfield  {journal} {\bibinfo  {journal} {Trends in
  Genetics},\ }\textbf {\bibinfo {volume} {25(8)}},\ \bibinfo {pages} {335}
  (\bibinfo {year} {2002})}\BibitemShut {NoStop}%
\bibitem [{\citenamefont {Giorgetti}\ \emph {et~al.}(2010)\citenamefont
  {Giorgetti}, \citenamefont {Siggers}, \citenamefont {Tiana}, \citenamefont
  {Caprara}, \citenamefont {Notarbartolo}, \citenamefont {Corona},
  \citenamefont {M.Pasparakis}, \citenamefont {Milani}, \citenamefont {Bulyk},\
  and\ \citenamefont {Natoli}}]{Giorgetti_Siggers_2010}%
  \BibitemOpen
  \bibfield  {author} {\bibinfo {author} {\bibfnamefont {L.}~\bibnamefont
  {Giorgetti}}, \bibinfo {author} {\bibfnamefont {T.}~\bibnamefont {Siggers}},
  \bibinfo {author} {\bibfnamefont {G.}~\bibnamefont {Tiana}}, \bibinfo
  {author} {\bibfnamefont {G.}~\bibnamefont {Caprara}}, \bibinfo {author}
  {\bibfnamefont {S.}~\bibnamefont {Notarbartolo}}, \bibinfo {author}
  {\bibfnamefont {T.}~\bibnamefont {Corona}}, \bibinfo {author} {\bibnamefont
  {M.Pasparakis}}, \bibinfo {author} {\bibfnamefont {P.}~\bibnamefont
  {Milani}}, \bibinfo {author} {\bibfnamefont {M.}~\bibnamefont {Bulyk}}, \
  and\ \bibinfo {author} {\bibfnamefont {G.}~\bibnamefont {Natoli}},\
  }\href@noop {} {\bibfield  {journal} {\bibinfo  {journal} {Mol. Cell},\
  }\textbf {\bibinfo {volume} {37(3)}},\ \bibinfo {pages} {418} (\bibinfo
  {year} {2010})}\BibitemShut {NoStop}%
\bibitem [{\citenamefont {Espinosa-Soto}\ \emph {et~al.}(2004)\citenamefont
  {Espinosa-Soto}, \citenamefont {Padilla-Longoria},\ and\ \citenamefont
  {Alvarez-Buylla}}]{Espinosa-Soto_Padilla-Longoria_2004}%
  \BibitemOpen
  \bibfield  {author} {\bibinfo {author} {\bibfnamefont {C.}~\bibnamefont
  {Espinosa-Soto}}, \bibinfo {author} {\bibfnamefont {P.}~\bibnamefont
  {Padilla-Longoria}}, \ and\ \bibinfo {author} {\bibfnamefont
  {E.}~\bibnamefont {Alvarez-Buylla}},\ }\href@noop {} {\bibfield  {journal}
  {\bibinfo  {journal} {The Plant Cell},\ }\textbf {\bibinfo {volume} {16}},\
  \bibinfo {pages} {2923} (\bibinfo {year} {2004})}\BibitemShut {NoStop}%
\bibitem [{\citenamefont {Chickarmane}\ and\ \citenamefont
  {Peterson}(2008)}]{Chickarmane_Peterson_2008}%
  \BibitemOpen
  \bibfield  {author} {\bibinfo {author} {\bibfnamefont {V.}~\bibnamefont
  {Chickarmane}}\ and\ \bibinfo {author} {\bibfnamefont {C.}~\bibnamefont
  {Peterson}},\ }\href@noop {} {\bibfield  {journal} {\bibinfo  {journal} {PLoS
  One},\ }\textbf {\bibinfo {volume} {3(10)}},\ \bibinfo {pages} {e3478}
  (\bibinfo {year} {2008})}\BibitemShut {NoStop}%
\bibitem [{\citenamefont {Davidich}\ and\ \citenamefont
  {Bornholdt}(2008)}]{Davidich_Bornholdt_2008}%
  \BibitemOpen
  \bibfield  {author} {\bibinfo {author} {\bibfnamefont {M.}~\bibnamefont
  {Davidich}}\ and\ \bibinfo {author} {\bibfnamefont {S.}~\bibnamefont
  {Bornholdt}},\ }\href@noop {} {\bibfield  {journal} {\bibinfo  {journal}
  {PLoS One},\ }\textbf {\bibinfo {volume} {3(2)}},\ \bibinfo {pages} {e1672}
  (\bibinfo {year} {2008})}\BibitemShut {NoStop}%
\bibitem [{\citenamefont {Ciliberti}\ \emph {et~al.}(2007)\citenamefont
  {Ciliberti}, \citenamefont {Martin},\ and\ \citenamefont
  {Wagner}}]{CilibertiMartin2007b}%
  \BibitemOpen
  \bibfield  {author} {\bibinfo {author} {\bibfnamefont {S.}~\bibnamefont
  {Ciliberti}}, \bibinfo {author} {\bibfnamefont {O.}~\bibnamefont {Martin}}, \
  and\ \bibinfo {author} {\bibfnamefont {A.}~\bibnamefont {Wagner}},\
  }\href@noop {} {\bibfield  {journal} {\bibinfo  {journal} {PLoS C.B.},\
  }\textbf {\bibinfo {volume} {3(2)}},\ \bibinfo {pages} {e15} (\bibinfo {year}
  {2007})}\BibitemShut {NoStop}%
\bibitem [{\citenamefont {Wagner}(2005)}]{WagnerBook2005}%
  \BibitemOpen
  \bibfield  {author} {\bibinfo {author} {\bibfnamefont {A.}~\bibnamefont
  {Wagner}},\ }\href@noop {} {\emph {\bibinfo {title} {Robustness and
  Evolvability in Living Systems}}}\ (\bibinfo  {publisher} {Princeton
  University Press},\ \bibinfo {address} {Princeton, NJ},\ \bibinfo {year}
  {2005})\BibitemShut {NoStop}%
\bibitem [{\citenamefont {Thieffry}\ \emph {et~al.}(1998)\citenamefont
  {Thieffry}, \citenamefont {Huerta}, \citenamefont {Perez-Rueda},\ and\
  \citenamefont {Collado-Vides}}]{ThieffryHuerta1998}%
  \BibitemOpen
  \bibfield  {author} {\bibinfo {author} {\bibfnamefont {D.}~\bibnamefont
  {Thieffry}}, \bibinfo {author} {\bibfnamefont {A.}~\bibnamefont {Huerta}},
  \bibinfo {author} {\bibfnamefont {E.}~\bibnamefont {Perez-Rueda}}, \ and\
  \bibinfo {author} {\bibfnamefont {J.}~\bibnamefont {Collado-Vides}},\
  }\href@noop {} {\bibfield  {journal} {\bibinfo  {journal} {Bio Essays},\
  }\textbf {\bibinfo {volume} {20}},\ \bibinfo {pages} {433} (\bibinfo {year}
  {1998})}\BibitemShut {NoStop}%
\bibitem [{\citenamefont {Balazsi}\ \emph {et~al.}(2008)\citenamefont
  {Balazsi}, \citenamefont {Heath}, \citenamefont {Shi},\ and\ \citenamefont
  {Gennaro}}]{BalazsiHeath2008}%
  \BibitemOpen
  \bibfield  {author} {\bibinfo {author} {\bibfnamefont {G.}~\bibnamefont
  {Balazsi}}, \bibinfo {author} {\bibfnamefont {A.}~\bibnamefont {Heath}},
  \bibinfo {author} {\bibfnamefont {L.}~\bibnamefont {Shi}}, \ and\ \bibinfo
  {author} {\bibfnamefont {M.}~\bibnamefont {Gennaro}},\ }\href@noop {}
  {\bibfield  {journal} {\bibinfo  {journal} {Mol. Syst. Biol.},\ }\textbf
  {\bibinfo {volume} {4}},\ \bibinfo {pages} {225} (\bibinfo {year}
  {2008})}\BibitemShut {NoStop}%
\bibitem [{\citenamefont {Frey}\ and\ \citenamefont
  {Dueck}(2007)}]{FreyDueck2007}%
  \BibitemOpen
  \bibfield  {author} {\bibinfo {author} {\bibfnamefont {B.}~\bibnamefont
  {Frey}}\ and\ \bibinfo {author} {\bibfnamefont {D.}~\bibnamefont {Dueck}},\
  }\href@noop {} {\bibfield  {journal} {\bibinfo  {journal} {Science},\
  }\textbf {\bibinfo {volume} {315}},\ \bibinfo {pages} {972} (\bibinfo {year}
  {2007})}\BibitemShut {NoStop}%
\bibitem [{\citenamefont {Milo}\ \emph {et~al.}(2002)\citenamefont {Milo},
  \citenamefont {Shen-Orr}, \citenamefont {Itzkovitz}, \citenamefont {Kashtan},
  \citenamefont {Chklovskii},\ and\ \citenamefont {Alon}}]{MiloShen-Orr2002}%
  \BibitemOpen
  \bibfield  {author} {\bibinfo {author} {\bibfnamefont {R.}~\bibnamefont
  {Milo}}, \bibinfo {author} {\bibfnamefont {S.}~\bibnamefont {Shen-Orr}},
  \bibinfo {author} {\bibfnamefont {S.}~\bibnamefont {Itzkovitz}}, \bibinfo
  {author} {\bibfnamefont {N.}~\bibnamefont {Kashtan}}, \bibinfo {author}
  {\bibfnamefont {D.}~\bibnamefont {Chklovskii}}, \ and\ \bibinfo {author}
  {\bibfnamefont {U.}~\bibnamefont {Alon}},\ }\href@noop {} {\bibfield
  {journal} {\bibinfo  {journal} {Science},\ }\textbf {\bibinfo {volume}
  {298}},\ \bibinfo {pages} {824} (\bibinfo {year} {2002})}\BibitemShut
  {NoStop}%
\bibitem [{\citenamefont {Maslov}\ and\ \citenamefont
  {Sneppen}(2002)}]{MaslovSneppen2002}%
  \BibitemOpen
  \bibfield  {author} {\bibinfo {author} {\bibfnamefont {S.}~\bibnamefont
  {Maslov}}\ and\ \bibinfo {author} {\bibfnamefont {K.}~\bibnamefont
  {Sneppen}},\ }\href@noop {} {\bibfield  {journal} {\bibinfo  {journal}
  {Science},\ }\textbf {\bibinfo {volume} {296}},\ \bibinfo {pages} {910}
  (\bibinfo {year} {2002})}\BibitemShut {NoStop}%
\bibitem [{\citenamefont {Barkai}\ and\ \citenamefont
  {Leibler}(2000)}]{BarkaiLeibler_2000}%
  \BibitemOpen
  \bibfield  {author} {\bibinfo {author} {\bibfnamefont {N.}~\bibnamefont
  {Barkai}}\ and\ \bibinfo {author} {\bibfnamefont {S.}~\bibnamefont
  {Leibler}},\ }\href@noop {} {\bibfield  {journal} {\bibinfo  {journal}
  {Nature},\ }\textbf {\bibinfo {volume} {403}},\ \bibinfo {pages} {267}
  (\bibinfo {year} {2000})}\BibitemShut {NoStop}%
\bibitem [{\citenamefont {Guantes}\ and\ \citenamefont
  {Poyatos}(2008)}]{GuantesPoyatos_2008}%
  \BibitemOpen
  \bibfield  {author} {\bibinfo {author} {\bibfnamefont {R.}~\bibnamefont
  {Guantes}}\ and\ \bibinfo {author} {\bibfnamefont {J.}~\bibnamefont
  {Poyatos}},\ }\href@noop {} {\bibfield  {journal} {\bibinfo  {journal} {PLoS
  C.B.},\ }\textbf {\bibinfo {volume} {4(11)}},\ \bibinfo {pages} {e1000235}
  (\bibinfo {year} {2008})}\BibitemShut {NoStop}%
\end{thebibliography}
%


\end{document}